\documentclass[nobibnotes,amsmath,amssymb,nofootinbib,notitlepage,prd,superscriptaddress,twocolumn]{revtex4-2}
\usepackage[T1]{fontenc}
\usepackage[usenames,dvipsnames]{xcolor}
\usepackage{graphicx,hyperref,orcidlink,float}
\hypersetup{colorlinks=true,citecolor=RoyalBlue,linkcolor=RoyalBlue,urlcolor=RoyalBlue}

\usepackage{appendix}
\usepackage{graphicx}

\usepackage{tikz}
\usepackage{pgfplots}
\usepackage{dcolumn}
\usepackage{bm}
\usepackage{amsmath}

\usepackage[mathlines]{lineno}
\usepackage{xcolor}
\usepackage{slashed}
\usepackage{booktabs}

\usepackage[commentmarkup=uwave,commandnameprefix=ifneeded]{changes}
\definechangesauthor[name=Harrison, color=teal]{HS}

\definechangesauthor[name=Keefe, color=orange]{KM}

\usepackage[normalem]{ulem}
\definechangesauthor[name=Hengrui, color=red]{HZ}

\definechangesauthor[name=Will, color=cyan]{WF}
\usepackage[normalem]{ulem}
\definechangesauthor[name=Max, color=magenta]{MI}

\usepackage[normalem]{ulem}

\newcommand{\dd}{\mathrm{d}}

\pgfplotsset{compat=1.18}
\begin{document}

\title{Black hole spectroscopy for precessing binary black hole coalescences}

\newcommand{\sbuaffil}{\affiliation{Department of Physics and Astronomy, Stony Brook University, Stony Brook NY 11794, USA}}
\newcommand{\ccaaffil}{\affiliation{Center for Computational Astrophysics, Flatiron Institute, New York NY 10010, USA}}
\newcommand{\cuaffil}{\affiliation{
 Department of Physics, Columbia University,
704 Pupin Hall, 538 West 120th Street, New York, New York 10027, USA
}}
\newcommand{\CornellPhysics}{\affiliation{Department of Physics, Cornell University, Ithaca,
  NY, 14853, USA}}
\newcommand{\Cornell}{\affiliation{Cornell Center for Astrophysics and Planetary
    Science, Cornell University, Ithaca, New York 14853, USA}}
\newcommand{\Caltech}{\affiliation{Theoretical Astrophysics, Walter Burke
  Institute for Theoretical Physics, California Institute of Technology,
  Pasadena, California 91125, USA}}
\newcommand{\Perimeter}{\affiliation{Perimeter Institute for Theoretical Physics, 31 Caroline Streeth North, Waterloo, Onatrio NSL 2Y5, Canada}}
\newcommand{\AEI}{\affiliation{Max Planck Institute for Gravitational Physics (Albert Einstein Institute), Am M\"uhlenberg 1, D-14476 Potsdam, Germany}}
\newcommand{\Dartmouth}{\affiliation{Department of Mathematics, Center for Scientific Computing and Data Science Research, University of Massachusetts, Dartmouth, MA 02747, USA}}

\author{Hengrui Zhu \orcidlink{0000-0001-9027-4184}}
\email{hengrui.zhu@princeton.edu}
\affiliation{Department of Physics, Princeton University, Jadwin Hall, Washington Road, New Jersey, 08544, USA}
\affiliation{Princeton Gravity Initiative, Princeton University, Princeton, New Jersey, 08544, USA}

\author{Harrison Siegel \orcidlink{0000-0001-5161-4617}}
\email{hs3152@columbia.edu}
\cuaffil
\ccaaffil
\author{Keefe Mitman \orcidlink{0000-0003-0276-3856}}
\email{kmitman@caltech.edu}
\Caltech
\author{Maximiliano Isi \orcidlink{0000-0001-8830-8672}}
\ccaaffil
\author{Will M. Farr \orcidlink{0000-0003-1540-8562}}
\ccaaffil
\sbuaffil
  
\author{\\Michael Boyle \orcidlink{0000-0002-5075-5116}}
\Cornell
\author{Nils Deppe \orcidlink{0000-0003-4557-4115}} \Cornell\CornellPhysics
\author{Lawrence E.~Kidder \orcidlink{0000-0001-5392-7342}} \Cornell
\author{Sizheng Ma \orcidlink{0000-0002-4645-453X}} \Perimeter
\author{Jordan Moxon \orcidlink{0000-0001-9009-6955}} \Caltech
\author{Kyle C. Nelli \orcidlink{0000-0003-2426-8768}} \Caltech
\author{Harald P. Pfeiffer \orcidlink{0000-0001-9288-519X}} \AEI
\author{Mark A. Scheel \orcidlink{0000-0001-6656-9134}} \Caltech
\author{Saul A. Teukolsky \orcidlink{0000-0001-9765-4526}} \Caltech\Cornell
\author{William Throwe \orcidlink{0000-0001-5059-4378}} \Cornell 
\author{Vijay Varma \orcidlink{0000-0002-9994-1761}} \Dartmouth
\author{Nils L. Vu \orcidlink{0000-0002-5767-3949}} \Caltech

\begin{abstract}
\noindent The spectroscopic study of black hole quasinormal modes in gravitational-wave ringdown observations is hindered by our ignorance of which modes should dominate astrophysical signals for different binary configurations, limiting tests of general relativity and astrophysics. In this work, we present a description of the quasinormal modes that are excited in the ringdowns of comparable mass, quasicircular precessing binary black hole coalescences---a key region of parameter space that has yet to be fully explored within the framework of black hole spectroscopy. We suggest that the remnant perturbation for precessing and nonprecessing systems is approximately the same up to a rotation, which implies that the relative amplitudes of the quasinormal modes in both systems are also related by a rotation. We present evidence for this by analyzing an extensive catalog of numerical relativity simulations. Additional structure in the amplitudes is connected to the system's kick velocity and other asymmetries in the orbital dynamics. We find that the ringdowns of precessing systems need not be dominated by the ${(\ell,m,n)=(2,\pm 2,0)}$ quasinormal modes, and that instead the $(2,\pm 1,0)$~or~$(2,0,0)$ quasinormal modes can dominate. Our results are consistent with a ringdown analysis of the LIGO-Virgo gravitational wave signal GW190521, and may also help in understanding phenomenological inspiral-merger-ringdown waveform model systematics.
\end{abstract}

\maketitle

\section{Introduction}
According to general relativity, the coalescence of two black holes results in the formation of a perturbed remnant, which equilibrates to a Kerr black hole by emitting gravitational waves, in a process called the ringdown~\cite{Teukolsky:1973ha, Buonanno:2006ui,Owen:2009sb}. Ringdown emission can in general consist of a transient burst that is succeeded by a spectrum of quasinormal modes (QNMs), which then themselves decay to reveal a power-law tail~\cite{Andersson_BHPerturbationSurvey1999, Frolov_BlackHolePhysicsTextbook}. The QNMs are exponentially damped sinusoids with complex frequencies that are uniquely determined by the mass and spin of the remnant black hole~\cite{Vishveshwara:1970cc,Press:1971wr,Teukolsky:1973ha,Chandrasekhar:1975zza,Leaver:1985ax, Detweiler:1980gk} and complex amplitudes that are nontrivially related to the initial conditions of the system~\cite{berti:2006kk, Leaver:1986gd}.

Much remains to be understood about how the progenitor properties of coalescing black holes map to the QNMs radiated by the remnant black hole. Numerical relativity (NR) studies of the ringdowns of comparable mass quasicircular binary black hole coalescences have largely been confined to nonprecessing systems, where the black hole spins are perpendicular to the orbital plane. These studies have shown empirically that some progenitor information, such as the mass ratio and spin magnitudes, is encoded in the QNM amplitudes~\cite{Kamaretsos:2012bs, Kamaretsos_BHHairLoss, London_ModelingRingdownII, Xiang_Ringdown, BorhanianRingdown, FortezaRingdown, London_modelingringdownbeyondfundqnms}.

For precessing systems~\cite{Taxonomy_precession, Apostolatos_Precession, Kidder_Precession1995PRD,Buonanno:2002fy}, where the progenitor black holes possess spin components parallel to the orbital plane, the picture becomes more complicated. In the extreme mass-ratio case, the relative amplitude ratios between QNMs are dictated by the spin magnitude of the heavier black hole and a small number of orbital parameters~\cite{Apte_EMRIprecessingringdown, Lim:2019xrb, Hughes:2019zmt, Ghosh_freqdomainPhenomPrecession}, which could be related to the connection between geodesics and QNMs in the eikonal limit \cite{Ferrari:1984zz,Mashhoon:1985cya,Dolan:2010wr,Yang:2012he,Hadar:2022xag}. For comparable mass systems, waveform models exist which map nonprecessing waveforms to precessing ones through a time-dependent, noninertial frame rotation to the coprecessing frame~\cite{Buonanno:2002fy,Schmidt:2012rh, OShaughnessy:2012iol,Boyle_GeometricApproach,Pan:2013rra}, and these mappings have been extended to ringdown waveform modeling in, e.g., Refs.~\cite{OShaughnessy:2012iol, Eleanor_RingdownFreqPrecessing, Babak:2016tgq, Ossokine:2020kjp, Ramos-Buades:2021adz}. Studies of the mismatch between NR waveforms and explicit QNM models, building on Ref.~\cite{Giesler:2019uxc}, have also been performed for precessing systems~\cite{Finch:PrecessionRingdown}.

In this work, we analyze the ringdowns of precessing black hole binaries and present a description of QNM excitations from the perspective of observational black hole spectroscopy and perturbation theory. We fit the amplitude of each individual QNM in the canonical frame for perturbation theory---the superrest frame of the remnant~\cite{Mitman:2021xkq,MaganaZertuche:2021syq,Mitman:2022kwt,Gualtieri:2008ux}. This means that any large amplitudes we recover will correspond to large contributions of the associated QNM frequencies in LIGO-Virgo-KAGRA~\cite{AdvancedLIGOScientific:2014pky,VIRGO:2014yos,KAGRA:2020tym} data, and correspond to a physically distinct state of the remnant after the merger. We analyze the ringdown within 250 NR simulations of both precessing and nonprecessing binary black hole coalescences~\cite{SXS_catalogue}.

We argue that the remnant perturbations in both precessing and nonprecessing systems share approximately the same geometry, but are rotated away from the $z$-axis with respect to each other. The angle of this rotation is determined by the misalignment of the total angular momentum flux $\dd\mathbf{J}/\dd t$ near the time of merger with the final spin $\boldsymbol{{\chi}}_f$ of the remnant black hole. Because precessing systems can also have asymmetric inspiral emission on either side of the orbital plane, there should also be further structure in the QNM amplitudes related to properties of the system's kick velocity~\cite{Ghosh_freqdomainPhenomPrecession}.

We show that in some precessing systems, unlike commonly assumed in spectroscopic studies, the dominant QNM will \textit{not} be an ${(\ell,m, n)=(2,\pm 2, 0)}$ QNM, but rather the $(2,\pm1, 0)$ or $(2,0, 0)$ QNMs. We also find that the precession-induced asymmetry in the gravitational wave emission can, in extreme cases, result in $\mathcal{O}(10)$ amplitude differences between $\pm m$ QNMs that share the same frequency.
Our findings have broad implications for the interpretation of ringdown signals in data from LIGO-Virgo-KAGRA and future detectors, and may also inform investigations into the nonlinear aspects of the ringdown. Furthermore, our work provides insight into how some systematic errors may arise in current phenomenological inspiral-merger-ringdown (IMR) waveform models~\cite{MacUilliam:2024oif, Dhani:2024jja}.

Our paper is organized as follows. In Sec.~\ref{sec:Methods}, we discuss our QNM fitting procedure, and define parameters quantifying precession dynamics which we make use of throughout the paper. In Sec.~\ref{sec:results}, we present our main results. We conclude in Sec.~\ref{sec:conclusion}. In the appendices, we discuss more technical aspects of our fitting procedure, and suggest an example of how our work may help to inform phenomenological IMR waveform modeling.

Our fitting and plotting codes, as well as an example simulation output, can be found in Ref.~\cite{GitHubRepo}.

\subsection{QNM conventions} Throughout this work, we will refer to individual QNMs by four indices $(p,\ell,m,n)$, following the notation in Ref.~\cite{Isi:2021iql}. The angular structure of the QNMs is described by spin-weighted spheroidal harmonics with angular indices $\ell$ and $m$. The radial structure of the QNMs is denoted by the index $n$, and is also tied to the lifetime of the QNMs: the longest-lived, ${n=0}$, QNMs are referred to as fundamental QNMs; and faster-decaying, ${n>0}$, QNMs are called overtones. For a given set of $(\ell,m,n)$, when ${m\neq0}$, there are two distinct QNMs which are labeled by an index $p\equiv\mathrm{sgn}[m\,\mathfrak{Re}\left(\omega\right)]$, where $\mathfrak{Re}\left(\omega\right)$ is the real part of the complex QNM frequency $\omega$, and whose phase fronts are either corotating (${p=+}$) or counterrotating (${p=-}$) with the black hole; solutions with ${m=0}$ are azimuthally symmetric, so that there is no notion of co vs counterrotating fronts and the two possible signs of ${\mathfrak{Re}\left(\omega\right)}$ directly encode the polarization degrees of freedom. We define $A_{(p,\ell,m,n)}$ as the norm of the complex amplitude for a given $(p,\ell,m,n)$ QNM at the time of peak luminosity.

\section{Methods}\label{sec:Methods}
\subsection{Extracting quasinormal modes}
We analyzed simulations of 226 precessing and 24 nonprecessing binary black hole coalescences, with mass ratios ${q\leq8}$ and component spin magnitudes ${\chi\leq0.8}$, created with the spectral Einstein code (\textsc{SpEC})~\cite{SXS_catalogue}. We computed the strain $h$ and the five Weyl scalars at future null infinity using the \textsc{SpECTRE} code's Cauchy-characteristic evolution (CCE) module~\cite{spectrecode,Moxon:2020gha,Moxon:2021gbv}. With this asymptotic data, we mapped the simulated systems to the superrest frame of their remnant black hole $250M$ past the peak of the strain's luminosity (where $M$ is the total Christodoulou mass of the binary), using the BMS frame fixing procedure outlined in Ref.~\cite{Mitman:2022kwt} and the code \textsc{scri}~\cite{mike_boyle_2020_4041972,WignerDQuaternions_Reference,Boyle_Precessing,Boyle:2015nqa}. Consequently, each simulation has been transformed such that its remnant black hole is at rest at the origin, has its spin in the positive-$\hat{z}$ direction, and has no Moreschi supermomentum~\cite{Mitman:2021xkq,Mitman:2022kwt,MaganaZertuche:2021syq}. This postprocessing is necessary in order to robustly extract physical QNM amplitudes \cite{MaganaZertuche:2021syq,Mitman:2022kwt}.

For each simulation, we simultaneously fit the strain using a QNM model with all fundamental modes $(p,\ell,m,0)$ for $p\in\{+,-\}$, $\ell\in\{2,3\}$, and $m\in\{-\ell,-\ell+1,...,\ell\}$. This model is fit over the two-sphere using a linear least-squares routine~\cite{Berti:2014fga,Xiang_Ringdown,Zhu:2023mzv} and the following procedure:\footnote{Similar fitting procedures have also been used in Refs.~\cite{London:2018gaq,Bhagwat:2019dtm,Baibhav:2023clw,Cheung:2023vki}.}
\begin{itemize}
    \item Fit the waveform from $t_{0}$ to $100M$ past $t_{\mathrm{peak}}$, where $t_{0}$ is the start time of the fit and is evenly varied from $20M$ to $80M$, in increments of $0.5M$, past $t_{\mathrm{peak}}$---note that $t_{\mathrm{peak}}$ is the time at which the strain's luminosity reaches its maximum value (see Eq.~(2.8) of Ref.~\cite{Ruiz:2007yx});
    \item Over a series of windows in $t_{0}$ with a length of $20M$, compute the fractional uncertainty for each of the QNM amplitudes;
    \item Compute the mean of these fractional uncertainties over the fitted QNMs;
    \item Check to see which $t_{0}$ window has the minimum mean fractional uncertainty and use this window to extract the amplitude (as a mean over the window) of each of the QNMs in the ringdown model.
\end{itemize}
This routine ensures that the QNMs are extracted over the window of start times in which their recovered amplitudes are all, on average, the most stable. We  do this rather than simply fitting at a fixed time because the timescales of the QNMs vary from system to system, depending on the magnitude of the remnant's spin. Over our entire catalog, this routine results in fits that have window start times ${t-t_{\mathrm{peak}}\approx33M\pm7M}$ and relative errors over the two-sphere of $\sim0.8\%$. Fitting at such late times enables us to avoid having to consider overtones and nonlinearities in our analysis. Nonetheless, for completeness we checked and confirmed that the inclusion of the first overtone for each QNM does not impact our results.

\subsection{Precession parameters}
In addition to the QNM amplitudes, we also compute two other precession-induced quantities in the superrest frame of the remnant: the angle $\theta$ between the total angular momentum flux $\dd\mathbf{J}/\dd t$ at $t_{\mathrm{peak}}$ and the final remnant spin $\boldsymbol{{\chi}}_f$, and the angle $\phi$ of the remnant kick velocity $\boldsymbol{v}$ with respect to $\boldsymbol{{\chi}}_f$. We formally define the remnant spin misalignment angle $\theta$ as
\begin{align}
    \label{eq:misalignmentangle}\theta=\mathrm{cos}^{-1}\left[\frac{\dd\mathbf{J}/\dd t\left(t=t_{\mathrm{peak}}\right)}{||\dd\mathbf{J}/\dd t\left(t=t_{\mathrm{peak}}\right)||}\cdot\frac{\boldsymbol{{\chi}}_f}{||\boldsymbol{{\chi}}_f||}\right],
\end{align}
where $\dd\mathbf{J}/\dd t$ is the total angular momentum flux (see Eq.~(2.24) of Ref.~\cite{Ruiz:2007yx}), $\boldsymbol{{\chi}}_f$ is the spin of the remnant black hole measured at future null infinity (see Eq.~(15) of Ref.~\cite{Iozzo:2021vnq}), and $||\cdot||$ is the $L^{2}$ norm. This angle shares similarities with the precession angles and optimal emission direction as identified in, e.g., Refs.~\cite{Eleanor_RingdownFreqPrecessing, Schmidt:2012rh, OShaughnessy:2012iol, Boyle_GeometricApproach} but, crucially, we will only use it to characterize theoretical predictions of the remnant perturbation's structure at the time of peak luminosity and not to track an effective ``coprecessing'' frame over time.

The kick angle $\phi$ is given by
\begin{align}
    \label{eq:velocityangle}\phi&=\cos^{-1}\left[\frac{\boldsymbol{v}_{\mathrm{CoM}}^{\mathrm{remnant}}-\boldsymbol{v}_{\mathrm{CoM}}^{\mathrm{binary}}}{||\boldsymbol{v}_{\mathrm{CoM}}^{\mathrm{remnant}}-\boldsymbol{v}_{\mathrm{CoM}}^{\mathrm{binary}}||}\cdot\frac{\boldsymbol{{\chi}}_f}{||\boldsymbol{{\chi}}_f||}\right]\nonumber\\
    &=\cos^{-1}\left[-\frac{\boldsymbol{v}_{\mathrm{CoM}}^{\mathrm{binary}}}{||\boldsymbol{v}_{\mathrm{CoM}}^{\mathrm{binary}}||}\cdot\frac{\boldsymbol{{\chi}}_f}{||\boldsymbol{{\chi}}_f||}\right],
\end{align}
where $\boldsymbol{v}_{\mathrm{CoM}}^{\mathrm{remnant}}=0$ in the superrest frame of the remnant, and $\boldsymbol{v}_{\mathrm{CoM}}^{\mathrm{binary}}$ is the premerger center-of-mass velocity (see Eq.~(12) of Ref.~\cite{Iozzo:2021vnq}) of the binary in the same frame. We extract the latter over a $500M$ window starting at $t=t_{\mathrm{peak}}-1000M$ using the procedure outlined in Ref.~\cite{Mitman:2022kwt}.

We make use of the Wigner-D matrices $\mathfrak{D}_{m^\prime,m}^{\ell}(\mathbf{R})$ that govern the mixing of spin-weighted spherical harmonics under rotations through the transformation \cite{WignerDQuaternions_Reference,Boyle_SWSH, wigner2012group, Schmidt_TrackingPrecession}
\begin{equation}
\label{eq:ampsum}
    \phantom{}_{s}Y_{(\ell,m')}' = \sum_m\mathfrak{D}_{m^\prime,m}^{\ell}\left(\mathbf{R}\right)\phantom{}_{s}Y_{(\ell,m)} \, ,
\end{equation}
where $\mathbf{R}$ is some quaternion, $\phantom{}_{s}Y_{(\ell,m)}$ is a spin-weight $s$ spherical harmonic, and primes indicate a rotated frame. The Wigner-D matrices mix amplitudes of harmonics with the same $\ell$ but different $m$.
Because we will be strictly interested in static rotations off the $\hat{z}$ axis by the remnant spin misalignment angle $\theta$, we simplify our notation to
\begin{align}
    \mathfrak{D}_{m^\prime,m}^{\ell}(\mathbf{R})\rightarrow \mathfrak{D}_{m^\prime,m}^{\ell}(\theta),
\end{align}
where now $\mathfrak{D}_{m^\prime,m}^{\ell}(\theta)$ represents a rotation by an angle $\theta$ off the $\hat{z}$ axis, but not about it. If we consider corotating and counterrotating perturbations, 
it can be shown that a fully corotating $+m$ perturbation, when rotated by ${\theta=\pi}$ off the remnant spin axis, would instead excite fully counterrotating $-m$ modes~\cite{PressTeukolsky_1973}.

\begin{figure*}[th]
   \includegraphics[width=\textwidth]{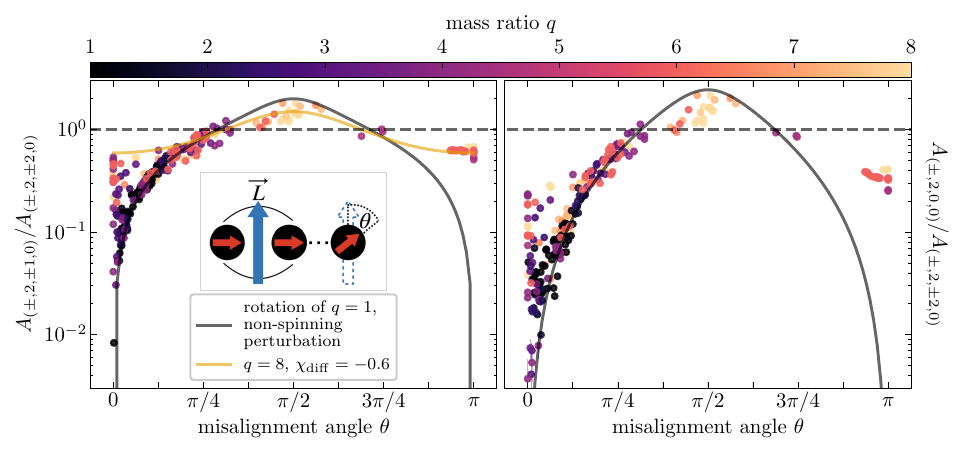}
\caption{\emph{Left}: Summed amplitude ratios defined, in Eq.~\eqref{eq:A+-def}, of the corotating and counterrotating $({\ell,m,n)=(2,\pm1,0)}$ and ${(\ell,m,n)=(2,\pm2,0)}$ QNMs, versus the remnant spin misalignment angle $\theta$, defined in Eq.~\eqref{eq:misalignmentangle}. Color shows the binary mass ratio. One can roughly predict the QNM amplitude ratio for comparable-mass precessing binaries by rotating an $({\ell,m)=(2,\pm2)}$ perturbation by $\theta$ off the $\hat{z}$ axis and taking the resulting ratio of different $(\ell, m)$ harmonics in the rotated frame. The black curve shows this prediction for ${q=1}$ systems, based on Eq.~\eqref{eq:rotpred}. For systems with a higher mass-ratio and nonzero spins, the ${\theta=0}$ perturbation to be rotated will intrinsically include significant ${\ell\not=m}$ content, dependent on the mass ratio and progenitor spins~\cite{Kamaretsos:2012bs, Kamaretsos_BHHairLoss,Berti:2005ys,Berti:2007fi,Berti:2007nw}. The yellow curve shows an example for the rotation of a ${q=8}$, ${\chi_{\mathrm{diff}}=-0.6}$ perturbation, per Eq.~\eqref{eq:A21A22}. Higher mass-ratio systems can achieve larger values of $\theta$, since the spin of the more massive black hole contributes a larger fraction of the remnant’s final spin in these systems. \emph{Right}: Identical to the other panel, but for the $(2,0,0)$ QNMs. Here we only show a prediction for a ${q=1}$, nonspinning system because models of the ${\theta=0}$ excitation of the $(2,0,0)$ QNM as a function of $q$ and $\chi_{\mathrm{diff}}$ are not as readily available as those of the $(2,1,0)$ QNM [cf.~Eq.~\eqref{eq:A21A22}]. Error bars are the standard deviation of the QNM amplitudes over the most stable window of fitting start times.
\label{fig:1stresult}}
\end{figure*}

\begin{figure}[t]
  \includegraphics[width=\columnwidth]{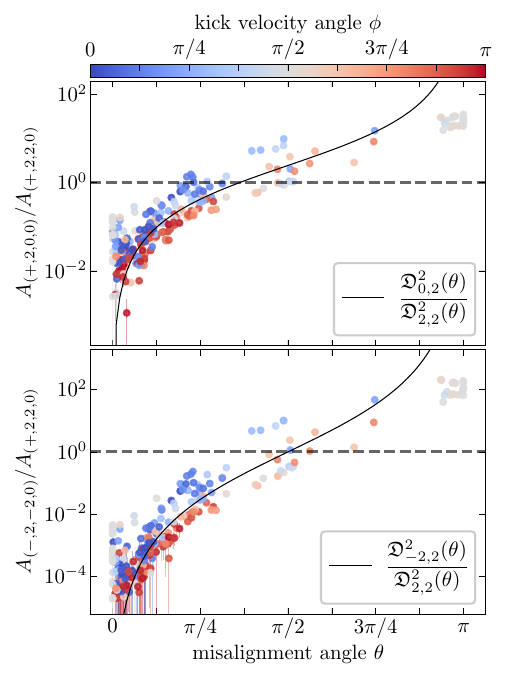}
  \caption{Top: amplitude ratio of the corotating $(+,2,0,0)$ and $(+,2,2,0)$ fundamental QNMs, versus the remnant spin misalignment angle $\theta$; see Eq.~\eqref{eq:misalignmentangle}. Color shows the angle between the kick velocity and the remnant spin direction; see Eq.~\eqref{eq:velocityangle}. The black curve is the prediction for this amplitude ratio based on the rotation of spin-weighted spherical harmonics assuming ${q=1}$ and a nonspinning binary. Bottom: identical to the top panel, but for the counterrotating $(-,2,-2,0)$ and corotating $(+,2,2,0)$ fundamental QNMs. The counterrotating QNMs dominate for $\theta\gtrsim\pi/2$ \cite{Eleanor_RingdownFreqPrecessing}. Fractional uncertainties here are larger than in the top panel, since the absolute values of the counterrotating amplitudes are close to zero for small values of $\theta$.}
  \label{fig:2ndresult}
\end{figure}

\section{Conceptual picture}

Before delving into our results, we first provide a conceptual framework that explains the ringdown amplitudes of precessing binary black hole coalescences by characterizing the structure of the remnant perturbations.

Evidence from NR suggests that the merger stage of a comparable-mass binary black hole coalescence is generally short-lived~\cite{Davis, Closelimit_PN,Buonanno:2006ui}. This implies that the state of the binary before merger may be directly related to the structure of the remnant perturbation immediately following merger~\cite{Kamaretsos:2012bs, Kamaretsos_BHHairLoss, BorhanianRingdown}.

In linear perturbation theory, the QNMs are sourced by perturbations to a fixed Kerr background corresponding to the final remnant. In keeping with this framework, we must study the structure of the perturbation sourced by the binary in a frame whose $z$ axis is aligned with the final remnant spin $\boldsymbol{{\chi}}_f$.

Because the dominant angular content of the inspiral emission in the coorbital frame for precessing systems is similar to that of nonprecessing systems (up to asymmetries over the orbital plane)~\cite{Boyle_Precessing, Boyle_GeometricApproach, Schmidt_TrackingPrecession, OShaughnessy:2011pmr, Schmidt:2012rh}, we assume the geometry (but not the orientation) of the remnant perturbation is roughly the same in both cases. We assume the total angular momentum of the perturbation is oriented along the binary orbital angular momentum $\mathbf{L}$ as evaluated at a time near merger, e.g., $t_{\rm peak}$. Because we do not have direct access to $\mathbf{L}$ in our simulations, in practice we use the total angular momentum flux $\dd\mathbf{J}/\dd t$ which is parallel to $\mathbf{L}$ up to 1 post-Newtonian (PN) order~\cite{Apostolatos_Precession,Kidder_Precession1995PRD}. 

In precessing systems, $\boldsymbol{{\chi}}_f$ will not necessarily be parallel to $\mathbf{L}$. This means that, as viewed by the remnant in the $\boldsymbol{{\chi}}_f$-aligned frame where QNMs are defined, the perturbation is rotated by the angle $\theta$ between $\mathbf{L}$ and $\boldsymbol{{\chi}}_f$. Therefore, the QNMs of a precessing system should be approximately sourced by a rotated version of the perturbation that acts on the remnant of a nonprecessing system, and the QNM amplitudes of precessing and nonprecessing systems should thus be related by a simple rotation of $\theta$ at a fixed time.

Given the above conceptual picture, we can obtain an approximation for the amplitude ratios of QNMs in precessing systems by rotating the angular content of the dominant perturbation expected in a nonprecessing ${\theta=0}$ system. For simplicity we assume, to reasonable approximation for equal mass nonspinning binaries, that the dominant ${\theta=0}$ perturbation is comprised solely of equal-amplitude corotating ${(\ell,m)=(2,\pm 2)}$ harmonics. We can also account for larger mass ratios or spins by additionally rotating odd $m$ perturbations~\cite{Kamaretsos:2012bs,Kamaretsos_BHHairLoss}. With the aforementioned assumptions, we predict that counterrotating QNM amplitudes are large for ${\theta > \pi/2}$, and ${\ell \neq m}$ QNM amplitudes are large for ${\theta \simeq \pi/2}$. As we will show below, these assumptions appear to accurately describe the behavior of precessing systems. In the case of highly spinning antialigned binaries, there may also be $\theta=0$ contributions from  ${(\ell,m)=(2,0)}$ perturbative content, since these binary configurations cause the final state to more closely resemble a head-on collision, and in principle these contributions should also be included in our rotated-perturbation prediction; we do not currently account for such perturbative content, and leave the study of this part of parameter space for future work.

As an example, we explicitly show here how we derive our prediction for the QNM amplitude ratio $A_{(\pm,2,\pm 1,0)}/A_{(\pm, 2,\pm 2,0)}$, where we have defined
\begin{align}
    A_{(\pm,\ell,\pm m,0)}^{2} &= A_{(+,\ell,m,0)}^{2}+A_{(-,\ell,m,0)}^{2} \nonumber\\
    &\phantom{=.}+A_{(+,\ell,-m,0)}^{2}+A_{(-,\ell,-m,0)}^{2}\, ,
    \label{eq:A+-def}
\end{align}
i.e., the norm of a multidimensional amplitude vector. Our rotated-perturbation prediction is given by
\begin{equation}\label{eq:rotpred}
    \frac{\mathfrak{D}_{1^\prime,2}^{\ell,\pm}(\theta,q,\chi_{\mathrm{diff}})}{\mathfrak{D}_{2^\prime,2}^{\ell,\pm}(\theta,q,\chi_{\mathrm{diff}})}~,
\end{equation}
where
\begin{equation}\label{eq:rotpred2}
    \begin{split}
        \left[\mathfrak{D}_{m_{1}^\prime,m_{2}}^{\ell,\pm}(\theta,q,\chi_{\mathrm{diff}})\right]^{2}&=\left[\frac{A_{(\ell,m_{1},0)}}{A_{(\ell,m_{2},0)}}\,\mathfrak{D}_{m_{1}^\prime,m_{1}}^{\ell,\pm}(\theta)\right]^{2}\\
        &\phantom{=.}+\left[\mathfrak{D}_{m_{1}^\prime,m_{2}}^{\ell,\pm}(\theta)\right]^{2}.
    \end{split}
\end{equation}
For the ratio of perturbative $A_{(+,2,1,0)}/A_{(+,2,2,0)}$ content to rotate by $\theta$ in Eq.~\eqref{eq:rotpred2}, we use the result of Ref~\cite{Kamaretsos:2012bs} for spin-aligned binaries, i.e.,
\begin{equation}\label{eq:A21A22}
    A_{(+,2,1,0)}/A_{(+,2,2,0)} \approx |0.43 (\sqrt{1-4\nu}-\chi_{\mathrm{diff}})|,
\end{equation}
where $\nu$ is the symmetric mass ratio,
\begin{equation}
    \nu = \frac{q}{(1+q)^2}
\end{equation}
and 
\begin{equation}
    \chi_{\mathrm{diff}} = \frac{q}{1+q}\chi_{1}-\frac{\chi_{1}+\chi_{2}}{2(1+q)}.
\end{equation}
One could also use the results of, e.g., Refs.~\cite{London_ModelingRingdownII,cheung2023extracting}; we use that of Ref.~\cite{Kamaretsos:2012bs} because it is simpler and nonetheless validates our argument. Note that in Eq.~\eqref{eq:rotpred2}
\begin{equation}
    \left[\mathfrak{D}_{m_{1}^\prime,m_{2}}^{\ell,\pm}(\theta)\right]^{2}=\left[\mathfrak{D}_{m_{1}^\prime,m_{2}}^{\ell}(\theta)\right]^{2}+\left[\mathfrak{D}_{-m_{1}^\prime,m_{2}}^{\ell}(\theta)\right]^{2}.
\end{equation}

\section{Results}\label{sec:results}


In Fig.~\ref{fig:1stresult}, we show that the rotated-perturbation prediction of Eq.~\eqref{eq:rotpred} tracks ${\ell=2}$ fundamental QNM amplitude ratios well as a function of $\theta$. Notably, we observe that for ${\theta \simeq \pi/2}$, the most dominant QNMs in comparable mass precessing systems will \textit{not} be the corotating or counterrotating $({\ell,m, n)=(2,\pm2,0)}$ QNMs, as is often assumed in spectroscopic tests, but rather the $(2,\pm 1,0)$ (left) or the $(2,0,0)$ QNMs (right). Even in cases where the $(2,\pm2,0)$ QNMs are intrinsically dominant when measured over the entire celestial sphere as we do here, precession-amplified ${\ell \neq m}$ QNMs can still appear dominant in observational data due to viewing angle effects and thus may play a significant role in observational black hole spectroscopy.


Despite overall good agreement, our data does have some spread around our theoretical prediction. Precessing systems exhibit asymmetries in their inspiral emission above and below the orbital plane. The asymmetries of the binary can be passed on to the perturbations of the remnant to create differences in the amplitudes of the $\pm m$ QNMs~\cite{Ghosh_freqdomainPhenomPrecession, London_modelingringdownbeyondfundqnms,Schnittman:2007ij}. This is partly responsible for the spread around our rotated-perturbation prediction in Fig.~\ref{fig:1stresult}. Asymmetric emission is associated with nonzero remnant kick velocities~\cite{GWmomentumtransport,Bruegmann_superkicks, Ma_superkick,Boyle:2015nqa,Alessio_2019, Bruegmann_superkicks,Pretorius:2007nq,Keppel:2009tc,Gerosa:2016vip}. In Fig.~\ref{fig:2ndresult} we show how the kick direction leaves imprints on the ringdown which neatly correspond to deviations from our rotated-perturbation predictions. It might be possible to model such effects by using 
additional PN information, as in Ref.~\cite{BorhanianRingdown} for example; we save this for future work. 

In the top panel of Fig.~\ref{fig:2ndresult} we compare the kick direction with spreads around our rotated-perturbation prediction for $A_{(+,2, 0,0)}/A_{(+, 2, 2,0)}$; examining only one sign of $m$ should make us more sensitive to hemispheric asymmetries. For $\theta\lesssim\pi/4$, where the $({\ell,m,n)=(2,\pm 2,0)}$ QNMs dominate, we find that the spread is indeed correlated with the kick velocity's direction: a kick in the positive-$\hat{z}$ direction corresponds to less radiated power in the $(2,2)$ mode, while a kick in the negative-$\hat{z}$ direction corresponds to more radiated power in the $(2,2)$ mode. However, when $\pi/4\lesssim\theta\lesssim3\pi/4$, then ${A_{(+,2,0,0)} \simeq A_{(+, 2,2,0)}}$, and the kick velocity is no longer as strongly correlated with only one $(\ell,\pm m, n)$ QNM.

In the bottom panel of Fig.~\ref{fig:2ndresult}, we show a similar result when comparing the counterrotating $(-,2,-2,0)$ and corotating $(+,2,2,0)$ QNM amplitudes; these amplitudes should be related as per the discussion in Sec.~\ref{sec:Methods}. When ${\theta > \pi/2}$, the counterrotating QNMs dominate over the corotating QNMs~\cite{Eleanor_RingdownFreqPrecessing}. Again, the spread of the amplitude ratio around the rotated-perturbation prediction is in part governed by the direction of the kick velocity with respect to the remnant spin.

\begin{figure}[t]
\includegraphics[width=0.496\textwidth]{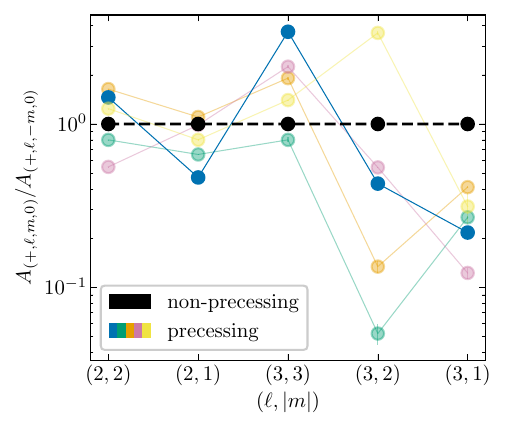}
  \caption{Amplitude ratio of corotating $\pm m$ QNMs, using a small subset of simulations in our catalog (see Table \ref{tab:sims}; order matches the order of the legend). Each color denotes a fit from an individual simulation. In precessing systems there are asymmetries in the emission power above and below the orbital plane, which are correlated with asymmetries in the QNM amplitudes over the remnant equatorial plane. The asymmetries for a given $\ell$ and $|m|$ harmonics can be large, $\mathcal{O}(10)$ in the extreme cases shown here, and distributed in diverse ways for different systems. Errors in the amplitude ratios are mostly too small to be seen.}
  \label{fig:dephased_asymm}
\end{figure}

The spread of systems near ${\theta = 0}$ in Fig.~\ref{fig:1stresult}, even for nearly equal mass systems, can also be partly attributed to asymmetric emission. Many of these systems are precessing despite having zero remnant spin misalignment, as shown in Fig.~\ref{fig:supp2} in Appendix~\ref{app:parity}. Most have progenitors whose spins are almost equal and opposite, meaning that the remnant spin direction is dominated by the direction of $\mathbf{L}$. These configurations can result in large kicks and consequently large asymmetries in the QNM excitations.

Not only can precession induce asymmetric excitations of QNMs, but because the phases and magnitudes of the precession-induced asymmetries in the inspiral are not the same for all $\ell$ and $m$ in the remnant frame, the QNM amplitude ratio $A_{(p,\ell,m,n)}/A_{(p,\ell,-m,n)}$ may vary for all $\ell$ and $|m|$ in a given system. PN expansions suggest that the magnitude of asymmetry in the dominant modes during the inspiral can become $\mathcal{O} (1)$ close to merger~\cite{Boyle_Precessing}; we show in Fig.~\ref{fig:dephased_asymm} that the degree of asymmetry in certain QNMs can be $\mathcal{O} (10)$ in the most extreme cases, and can be distributed in a variety of ways, presumably based on the orbital phase and the particular spin configuration of each system. Note that these extreme cases of asymmetry emerge in systems which are astrophysically plausible. As a result of emission asymmetries, precessing systems can produce radically different observable QNM amplitude spectra depending on the viewing angle of any terrestrial detectors.

\begin{figure*}[ht]
  \includegraphics[width=\textwidth]{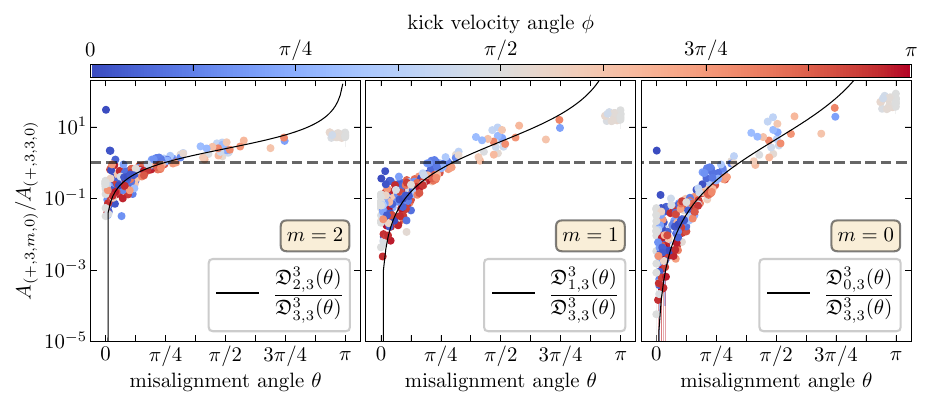}
  \caption{Amplitude ratio between corotating $(3,2,0)$, $(3,1,0)$, $(3,0,0)$, and the $(3,3,0)$ QNMs. The rotated-perturbation predictions (black curves) are obtained by rotating a $(3,\pm 3)$ spherical harmonic by $\theta$, i.e., by implicitly assuming that the dominant perturbation for ${\theta=0}$ consists solely of corotating $(3,\pm 3)$ content. Because our catalog mostly contains $q\neq 1$ precessing simulations, this assumption is reasonable. The kick velocity is not as straightforwardly correlated with ${\ell = 3}$ QNM amplitudes as it is with ${\ell = 2}$ QNM amplitudes (see Fig.~\ref{fig:2ndresult}).}
  \label{fig:lequals3rotations}
\end{figure*}

Lastly, we demonstrate that our conceptual picture seems to naturally extend to ${\ell=3}$ modes, and likely applies to higher $\ell$ as well. For nonequal mass nonprecessing systems, the $(3,3)$ mode typically dominates over the $(3,2)$ mode~\cite{Kamaretsos:2012bs}; thus we will assume that we can predict the $\ell=3$ QNM amplitudes of precessing systems by rotating a $(3,\pm 3)$ corotating perturbation by $\theta$. In Fig.~\ref{fig:lequals3rotations}, we show that the relative amplitudes between the corotating $\ell=3$, $m<3$ fundamental QNMs and the $(+,3,3,0)$ QNM follow this prediction. However, the spread around the prediction is no longer as clearly explained by the kick velocity angle $\phi$, because these modes are not as centrally tied to the kick direction due to their subdominant amplitudes.

\begin{table}[]
\caption{Parameters for simulations in Fig.~\ref{fig:dephased_asymm}}
\label{tab:sims}
    \centering
    \begin{tabular}{cccc}
    \toprule
    ID~\cite{GitHubRepo} & $q$ & $\vec{\chi}_{1}$ & $\vec{\chi}_{2}$ \\ [0.5ex] 
    \midrule
    $115$ & $1.5$ & $(-0.55, +0.41, -0.41)$ & $(+0.29, -0.32, -0.48)$ \\
    $124$ & $1.2$ & $(+0.14, -0.51, +0.39)$ & $(+0.32, +0.67, -0.21)$\\
    $96$ & $1.0$ & $(-0.73, -0.34, +0.01)$ & $(-0.35, -0.38, +0.61)$\\
    $97$ & $1.0$ & $(+0.07, +0.80, -0.01)$ & $(-0.29, -0.45, +0.59)$\\
    $162$ & $3.4$ & $(+0.19, -0.11, +0.75)$ & $(+0.33, +0.60, -0.12)$\\
    \bottomrule
    \end{tabular}
\end{table}

\section{Conclusion}\label{sec:conclusion}

By fitting the first-order fundamental quasinormal mode amplitudes in the ringdowns of 250 numerical relativity simulations of both nonprecessing and precessing binary black hole coalescences, we demonstrate in Figs.~\ref{fig:1stresult}~and~\ref{fig:2ndresult} that the ${\ell=2}$ remnant perturbations in precessing systems are related to those of nonprecessing systems by a rotation away from the $z$-axis. By this, we mean that the initial conditions which determine the loudness of ringdown frequencies in precessing and nonprecessing systems are approximately related in a straightforward way. We characterize this time-independent rotation by the misalignment of two quantities: the total angular momentum flux $\dd\mathbf{J}/\dd t$ near the time of merger and the final spin $\boldsymbol{{\chi}}_f$ of the remnant black hole. In Fig.~\ref{fig:lequals3rotations} we demonstrate that our framework for characterizing the remnant perturbation also extends to ${\ell=3}$.

Figs.~\ref{fig:1stresult}~and~\ref{fig:2ndresult} show that in some precessing systems the dominant quasinormal mode will \textit{not} be an ${(\ell,m,n)=(2,\pm 2,0)}$ quasinormal mode, but rather an ${(\ell,m,n)=(2,\pm 1,0)}$ or an ${(\ell,m)=(2,0,0)}$ quasinormal mode, contrary to what is commonly assumed in ringdown analyses of LIGO-Virgo-KAGRA data \cite{Carullo_GW150914, Isi_GW150914NoHair, 
Cotesta:2022pci,Crisostomi:2023tle,Gennari:2023gmx,Correia:2023bfn,PhysRevLett.131.169001,PhysRevLett.131.169002,Finch:2022ynt,Wang:2023xsy,Ma:2023vvr,Wang:2023mst,Isi:2022mhy,CalderonBustillo:2020rmh,
GW190521_Properties, Capano_GW190521, TestsGR_2ndCatLIGO}.

Additionally, we show in Fig.~\ref{fig:dephased_asymm} that precession-induced emission asymmetries, which are correlated with remnant kicks and other orbital dynamics, are encoded in the quasinormal mode spectra. These asymmetries can be large, such that amplitude ratios of individual $\pm m$ quasinormal modes which share the same frequency can be $\mathcal{O}(10)$ in some precessing systems. This implies that, for real measurements, the viewing angle can alter the observed quasinormal mode amplitudes of a precessing system more than would be possible in a nonprecessing system.

Crucially, the excitation of ${\ell\neq |m|, n}$ QNMs does not arise as an artifact from operating in a noninertial frame, as our fitting is performed in the superrest frame of the remnant, i.e., the inertial frame in which quasinormal modes are defined by perturbation theory (see Sec.~\ref{sec:Methods}). Our fits return the amplitude attached to the specific frequency of a physical $(p,\ell,m,n)$ quasinormal mode. This means that any large amplitudes we recover will correspond to large contributions of those associated frequencies in observed LIGO-Virgo-KAGRA data, which must be taken into account in order to accurately perform black hole spectroscopy.

Our results are consistent with a recent analysis~\cite{Siegel_GW190521} of the LIGO-Virgo gravitational wave signal GW190521\_030229~\cite{GW190521g_DiscoveryPaper}, wherein ringdown models with a large ${(\ell, |m|,n)=(2,1,0)}$ fundamental quasinormal mode amplitude were found to produce remnant mass and spin estimates in agreement with those of the \textsc{NRSur7dq4} waveform~\cite{NRSur7dq4_Paper}.\footnote{An alternative interpretation of this signal was originally put forth in Ref.~\cite{Capano_GW190521}, and replicated and commented on at length in Ref. ~\cite{Siegel_GW190521}.} Furthermore, our results are similar to Ref.~\cite{Hughes:2019zmt}, where it was found that the excitation of quasinormal modes in extreme mass-ratio systems is correlated with the polar angle of plunge of the smaller mass. The conceptual picture presented in our work may be valid over all binary mass ratios.

Our work may also help improve understanding of systematic biases in phenomenological inspiral-merger-ringdown waveforms~\cite{MacUilliam:2024oif, Dhani:2024jja}. As one example, we discuss in App.~\ref{app:phenom_waveforms} how our work highlights where a common assumption~\cite{OShaughnessy:2012iol} regarding the coprecessing frame optimal emission direction in these waveform models may break down. We also note that a model which explicitly fits QNMs, such as Ref.~\cite{SEOBNRv3}, might benefit from incorporating our simple rotated-perturbation prediction for the individual QNM amplitudes.

One might think that at times immediately following the peak strain, the rapidly changing emission direction of precessing systems might preclude the application of perturbation theory~\cite{Hamilton_PhenomPNR}. Our work suggests however that, at least to linear order in perturbation theory, information from the point of peak luminosity imprints itself on the ringdown of a precessing system in a straightforward way.

\textit{Software:} \textsc{seaborn}~\cite{Seaborn}, \textsc{matplotlib}~\cite{matplotlib}, \textsc{jupyter}~\cite{jupyter}, \textsc{numpy}~\cite{numpy}, \textsc{Python3}~\cite{Python3}.

\section*{ACKNOWLEDGMENTS}\label{sec:acknowledgements}

We thank Katerina Chatziioannou, Scott Hughes, Yuri Levin, Frans Pretorius, and Aaron Zimmerman for valuable conversations. We are also grateful to Eleanor Hamilton, Mark Hannam, Sascha Husa, Lionel London, Geraint Pratten, and Antoni Ramos-Buades for their insight on this work in the context of gravitational waveform modeling. Computations for this work were performed with the Wheeler cluster at Caltech. This work was supported in part by the Sherman Fairchild Foundation and NSF Grants No. PHY-2011968, PHY-2011961, PHY-2309211,  PHY-2309231, OAC-2209656 at Caltech., as well as NSF Grants No. PHY-2207342 and OAC-2209655 at Cornell. V.V.~acknowledges support from NSF Grant No. PHY-2309301.
The Flatiron Institute is a division of the Simons Foundation. H.S.'s research is supported by Yuri Levin's Simons Investigator Award 827103.


\appendix
\section*{Appendices}
\begin{figure}[h]
  \includegraphics[width=0.482\textwidth]{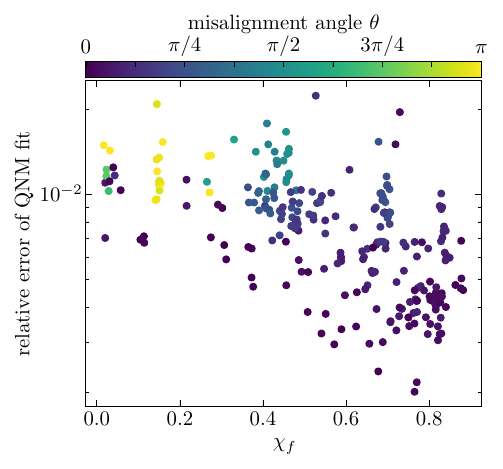}
  \caption{Relative error of QNM fit, versus the remnant spin, colored by the misalignment angle $\theta$. We find that the relative error decreases with increasing spin, as the QNMs are longer-lived. Precessing binaries in our catalog in general yield remnant spins smaller than the nonprecessing systems, thereby resulting in higher error, but the relative error remains on the order of 1\%.}
  \label{fig:supp1}
\end{figure}

\begin{figure}[h]
  \includegraphics[width=0.482\textwidth]{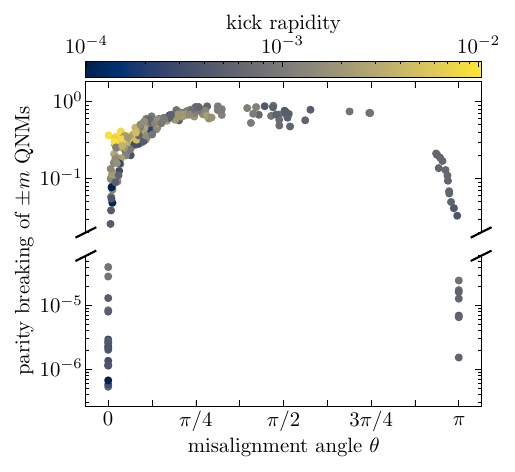}
  \caption{Parity breaking of $\pm m$ QNMs over the remnant equator as computed via Eq.~\eqref{eq:asymmetry}, versus the misalignment angle $\theta$. Nonprecessing systems maintain reflection symmetry over the remnant equator in the ringdown, whereas precessing systems do not: here the events with small asymmetries (below the vertical axis break) are nonprecessing. Note that many precessing systems have ${\theta\approx0}$ or $\pi$, i.e., remnant spin misalignment is not necessitated by precession. Especially at ${\theta\approx0}$, many precessing systems are characterized by large kick rapidities: many of these systems have roughly equal component spins fully antialigned with each other~\cite{Ma_superkick}, and thus the total angular momentum of the binary and subsequently the spin of the remnant is dominated by $\mathbf{L}$.}
  \label{fig:supp2}
\end{figure}

\section{ERROR OF QNM FITS}
\label{app:QNMfiterrors}
To characterize the quality of the QNM fits, we define the following fitting error:
\begin{equation}
    E(t_{0}) \equiv \sum_{\ell\in\{2,3\}}\sum_{|m|\leq \ell}\frac{\int_{t_0}^{100 M}\left\vert h_{(\ell,m)}-h_{(\ell,m)}^{\mathrm{QNM model}}\right\vert^2 dt}{\int_{t_0}^{100 M}\left\vert h_{(\ell,m)}\right\vert^2 dt}
\end{equation}
In Fig.~\ref{fig:supp1}, we plot the mean of $E(t_{0})$ over the most stable window of each simulation versus the magnitude of the remnant's spin, colored by the misalignment angle $\theta$. As can be seen, the fitting error is on the order of 1\% for every simulation.
Furthermore, we find that fitting error decreases with the magnitude of the remnant's spin. This is likely because the QNMs decay faster as the spin magnitude decreases, which implies that more of the fit to the numerical waveform is influenced by numerical noise.
As for the connection to the misalignment angle $\theta$, precessing black hole binaries require less fine-tuning of their spin configurations to produce slowlyspinning remnants, since the spin of the progenitor black holes need not be aligned with the angular orbital momentum. In the future, it could be useful to perform fits which have a varying end time based on the magnitude of the remnant's spin, but we save this for future work.

\section{PARITY BREAKING OVER THE REMNANT EQUATOR}
\label{app:parity}
One way to quantify the asymmetry of the $\pm m$ QNMs is by looking at their behavior under the $z$-parity operation, which is defined to act on the strain such that
\begin{equation}
    P_z(h_{(\ell, m)})=(-1)^\ell~\overline{h}_{(\ell,-m)},
\end{equation}
where the overbar denotes a complex conjugate. We can quantify the overall parity violation in the superrest frame by computing
\begin{align}
    \label{eq:asymmetry}
    P=\sqrt{\frac{\sum_{\ell,m}|h_{(\ell,m)}-(-1)^{\ell}~\overline{h}_{(\ell,-m)}|^{2}}{4\sum_{\ell,m}|h_{(\ell,m)}|^{2}}},
\end{align}
as described in~\cite{Boyle_Precessing}. For nonprecessing simulations, $P$ will be zero, up to numerical noise. However, precessing systems can exhibit significant parity breaking in the $\pm m$ QNMs.

In Fig.~\ref{fig:supp2} we show the parity violation of all the simulations in our catalog. The nonprecessing systems are clearly separated from the precessing systems, as denoted by the vertical axis break. Note that there are many precessing systems which have a remnant spin misalignment angle ${\theta\approx 0}$ or $\pi$. Remnant spin misalignment is not strictly required in precessing systems: fine-tuning of the spins, such as is found in superkick configurations~\cite{Ma_superkick} for example, can allow for precessing systems with ${\theta\approx 0}$, and large mass ratio systems can have essentially arbitrary $\theta$. \\

\section{IMPLICATIONS FOR PHENOMENOLOGICAL WAVEFORM MODELS}\label{app:phenom_waveforms}
Our work may provide a novel way of interpreting systematic biases associated with the ringdown stages of current phenomenological IMR waveform models. As one example, we consider here the way in which some phenomenological waveform models model the coprecessing frame of the ringdown, and the breakdown of an associated approximation for the time evolution of this frame.

Phenomenological IMR waveform models of precessing binary black holes model the ringdown by utilizing a noninertial ``coprecessing frame'' in which the angular harmonic mode decomposition is simplified~\cite{OShaughnessy:2011pmr, Schmidt_TrackingPrecession, Schmidt:2012rh}. In particular, these models take the angular harmonic mode content in the coprecessing frame to look like a nonprecessing system and then twist this content into an inertial frame via some time-dependent, noninertial rotation . During ringdown, some of these models track this rotation via the coprecessing frame’s Euler angles $(\alpha, \beta, \gamma)$, using an ansatz for $\dot{\alpha}$ and $\dot{\beta}$~\cite{OShaughnessy:2012iol}:
\begin{align}\label{eqn:oshaughnessy_ansatz}
\begin{split}
    \dot{\alpha} &= \mathfrak{Re}(\omega_{220} - \omega_{210}), \\
    \dot{\beta} &= \mathfrak{Im}(\omega_{220} - \omega_{210}),
\end{split}
\end{align}
where $\omega_{lmn}$ are complex frequencies associated with corotating QNMs. If the misalignment angle $\theta$ (or equivalently $\beta$) exceeds $\pi/2$, counterrotating QNM frequencies are used instead. For $\dot{\gamma}$, some models also adopt the minimal rotation condition~\cite{Boyle_GeometricApproach}:
\begin{equation}
     \dot{\gamma} = -\dot{\alpha}\cos(\beta).
     \label{eq:min_rot_cond}
\end{equation}




\begin{figure}[t]    \includegraphics[width=0.496\textwidth]{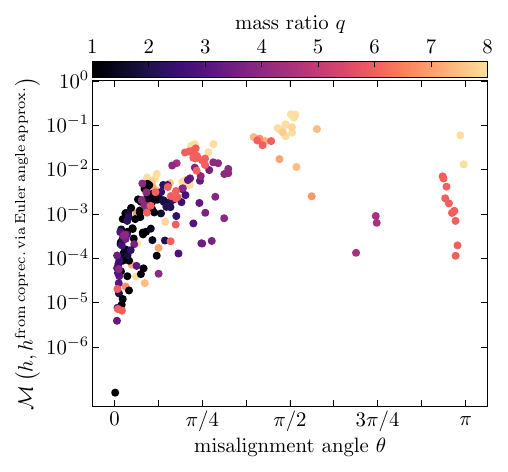}
  \caption{Systematic error associated with the coprecessing frame prescription for the ringdown Euler angles of Eq.~\eqref{eqn:oshaughnessy_ansatz}. This prescription is incorporated in some phenomenological IMR waveform models, and breaks down if the ${(\ell,m,n) = (2,0,0)}$ QNMs are excited. As we show in Fig.~\ref{fig:1stresult}, the $(2,0,0)$ QNMs are excited in many precessing configurations, and thus Eq.~\eqref{eqn:oshaughnessy_ansatz} may be a source of systematic error that affects current and future data analyses. On the horizontal axis we show the misalignment angle $\theta$. On the vertical axis, we show the mismatch (see Eq.~\eqref{eq:mismatch}) between the inertial frame waveform and a waveform that has been mapped from the coprecessing frame to the inertial frame using Eqs.~\eqref{eqn:oshaughnessy_ansatz} and~\eqref{eq:min_rot_cond}. Only precessing systems are shown. As the misalignment angle increases and the $(2,0)$ modes become relatively more excited, the mismatch increases due to the inaccuracy of Eq.~\eqref{eqn:oshaughnessy_ansatz}. Note that for $\theta\geq\pi/2$, we replace the corotating frequencies in Eq.~\eqref{eqn:oshaughnessy_ansatz} with counterrotating frequencies which have opposite-sign real components.}
  \label{fig:OShaughnessy_violation}
\end{figure}

As discussed in Ref.~\cite{OShaughnessy:2012iol} in its text around Eq.~(20), Eq.~\eqref{eqn:oshaughnessy_ansatz} only is a good approximation when the amplitude of the $(2,0,0)$ QNM in the remnant frame (i.e., the superrest frame) is zero.
Additionally, Eq.~\eqref{eqn:oshaughnessy_ansatz} assumes monotonic evolution of the Euler angles, though oscillations in the angles may occur due to the angular structure of the spheroidal harmonics and the beating between different modes (for evidence of this beating, see, e.g., Fig.~14 of Ref.~\cite{Marsat:2018oam}).

In Fig.~\ref{fig:OShaughnessy_violation}, we provide empirical evidence of systematic bias from Eq.~\eqref{eqn:oshaughnessy_ansatz}. The vertical axis shows the mismatch $\mathcal{M}$ between ringdown waveforms in the inertial frame of the remnant; the second waveform, however, has been mapped to this frame by mapping from the coprecessing frame to the inertial frame via the Euler angles presented in Eqs.~\eqref{eqn:oshaughnessy_ansatz} and~\eqref{eq:min_rot_cond} over the interval $[0M,100M]$. Consequently, the first waveform is the exact solution while the second waveform represents a waveform created in a similar fashion to phenomenological models. All $\ell\leq8$ modes are used for producing the data shown in Fig.~\ref{fig:OShaughnessy_violation}.

The ringdown mismatch is defined as:
\begin{align}
    \label{eq:mismatch}
    \mathcal{M}\left(h_{1},h_{2}\right)|_{t_{0}}^{t_{f}}\equiv1-\frac{O(h_{1},h_{2})}{\sqrt{O(h_{1},h_{1})O(h_{2},h_{2})}},
\end{align}
where
\begin{align}
    O(h_{1},h_{2})\equiv\int_{t_{0}}^{t_{f}}\int_{S^{2}}h_{1}\overline{h_{2}}\,d\Omega\,dt,
\end{align}
and $t_{0}=0M$ and $t_{f}=100M$. Higher mismatches indicate more of an inability of Eq.~\eqref{eqn:oshaughnessy_ansatz} to accurately track the true coprecessing frame. It can be seen that as the misalignment angle $\theta$ increases, the mismatch rises, reaching $\mathcal{O}(10^{-1})$ in regions where the $(2,0,0)$ QNMs dominate. Systems with $\theta\sim\pi$ also tend to have high mismatches due to their unequal mass ratios, which enable more refined polarizations and excitation of additional modes, complicating the optimal emission direction's time-dependence.

To our knowledge, Eq.~\eqref{eqn:oshaughnessy_ansatz} is used explicitly in the following IMR waveform models, although we note that this list may not be exhaustive:
\begin{itemize}
    \item TEOBResumS: Eq.~(21) in Ref.~\cite{Gamba:2021ydi}
    \item SEOBNRv4PHM: Eq.~(3.4) in Ref.~\cite{SEOBNRv4HM_Paper}
    \item SEOBNRv5PHM: Eqs.~(18--22) in Ref.~\cite{SEOBNRV5PHM}, via the term $\omega_\text{prec}$
    \item IMRPhenomTPHM: Eq.~(24) in Ref.~\cite{IMRPhenomTPHM}
\end{itemize}
Eq.~\eqref{eqn:oshaughnessy_ansatz} also appears either explicitly or implicitly in the following works: IMRPhenomXO4A~\cite{PhenomXO4a}, IMRPhenomPNR~\cite{Hamilton_PhenomPNR}, Ref.~\cite{Eleanor_RingdownFreqPrecessing} (through Eqs.~(9--10), Eqs.~(13--15), and Eq.~(35), although they also make use of NR tuning to the frequencies of some of the dominant modes), and Ref.~\cite{Marsat:2018oam}. Generally, when there are multiple options implemented for describing the ringdown coprecessing frame Euler angles $\alpha$ and $\beta$ in one of the above models, Eq.~\eqref{eqn:oshaughnessy_ansatz} is the default.

A model that explicitly fits QNMs, such as Ref.~\cite{SEOBNRv3}, might benefit from incorporating our simple rotated-perturbation prediction for the individual QNM amplitudes.


\bibliography{thebib}
\end{document}